# Graphdiyne: a two-dimensional thermoelectric material with high figure of merit

L. Sun, P. H. Jiang, H. J. Liu[*], D. D. Fan, J. H. Liang, J. Wei, L. Cheng, J. Zhang, J. Shi[†]

*Key Laboratory of Artificial Micro- and Nano-Structures of Ministry of Education and School of Physics and Technology, Wuhan University, Wuhan 430072, China*

As a new carbon allotrope, the recently fabricated graphdiyne has attracted much attention due to its interesting two-dimensional character. Here we demonstrate by multiscale computations that, unlike graphene, graphdiyne has a natural band gap, and simultaneously possess high electrical conductivity, large Seebeck coefficient, and low thermal conductivity. At a carrier concentration of $2.74 \times 10^{11}$ cm$^{-2}$ for holes and $1.62 \times 10^{11}$ cm$^{-2}$ for electrons, the room temperature $ZT$ value of graphdiyne can be optimized to 3.0 and 4.8, respectively, which makes it an ideal system to realize the concept of "phonon-glass and electron-crystal" in the thermoelectric community.

---

[*] Corresponding author. E-mail: phlhj@whu.edu.cn (H. J. Liu)
[†] Corresponding author. E-mail: jshi@whu.edu.cn (J. Shi)

## 1. Introduction

In the past several decades, plenty of efforts have been devoted to the fabrication and investigation of novel carbon allotropes, such as fullerene [1], carbon nanotube [2], and graphene [3]. In 2010, a new form of carbon allotrope named graphdiyne (GDY) was successfully synthesized on the surface of copper via cross coupling reaction [4]. GDY is a one-atom-thick two-dimensional material composed of $sp$ and $sp^2$ hybridized carbon atoms, where adjacent carbon hexagons are connected by two acetylenic linkages ($-C \equiv C-$) [5]. Unlike graphene with a Dirac cone-like electronic structure, GDY has a natural band gap of 0.47 eV [6]. Due to its unique two-dimensional structure, GDY is predicted to be the most stable diacetylenic carbon allotropes with high thermal resistance, high electrical conductivity, extreme hardness, and it is synthetically approachable [7, 8]. Many potential applications of GDY have been proposed, including gas separation [9], lithium storage [10], catalyst for dehydrogenation [11], and replacement for the existing silicon transistor [12].

The high thermal resistance together with high electrical conductivity of GDY is reminiscent of thermoelectric materials, which can directly convert heat into electricity and vice versa. The efficiency of a thermoelectric material is determined by its figure of merit or the $ZT=S^2\sigma T/(\kappa_e+\kappa_l)$, where larger electrical conductivity $\sigma$ and Seebeck coefficient $S$ along with smaller thermal conductivity (including both electronic part $\kappa_e$ and lattice part $\kappa_l$) are required for better performance. An ideal thermoelectric material behaves as glass for phonons and crystal for electrons (PGEC), as first proposed by Slack [13]. To be competitive with the traditional energy conversion methods, the *ZT* value of a thermoelectric material should be larger than 3.0. However, such target is still far from been reached. In this work, we combine first-principles calculations, Boltzmann theory, and molecular dynamics simulations to investigate the thermoelectric properties of GDY, and demonstrate that the room temperature *ZT* value of this novel two-dimensional system can be optimized to 3.0 and 4.8 for the *p*-type and *n*-type carriers, respectively. Such high *ZT* values significantly exceed most laboratory results reported so far, making GDY a plausible candidate for high-performance thermoelectric materials.

## 2. Computational method

The structure optimization and electronic properties of GDY are calculated by using the projector augmented wave (PAW) method within the framework of density functional theory (DFT). The code is implemented in the Vienna *ab-initio* Simulation Package [14, 15]. We use the generalized gradient approximation (GGA) with the exchange-correlation energy in the form of Perdew-Burke-Ernzerhof (PBE) [16]. The cutoff energy of 400 eV is chosen for the plane-wave expansion, and the Brillouin zones are sampled with a $11 \times 11 \times 1$ Monkhorst-Pack **k**-mesh. We adopt a standard supercell geometry where the GDY and its periodic images are aligned in a hexagonal array with large vacuum distance of 12 Å so that they can be treated as independent entities. The atomic positions of the GDY are fully relaxed until the magnitude of the forces acting on all the atoms becomes less than 0.05 eV/Å. The electronic transport coefficients ($S$, $\sigma$) are evaluated by using the Boltzmann theory [17], where the relaxation time is estimated from the deformation potential (DP) theory [18]. For the phonon transport, we use the equilibrium molecular dynamics (EMD) simulations, where the carbon-carbon interactions are described by the adaptive intermolecular reactive empirical bond order (AIREBO) potential [19] and the time step is set to 0.1 fs. The system is first simulated in an *NVT* (constant number of atoms, volume, and temperature) ensemble for 1,000,000 steps, and then switched into an *NVE* (constant number of atoms, volume, and energy) ensemble for 500,000 steps to reach the equilibrium state. The heat current data are obtained from another 15,000,000 steps in the *NVE* ensemble and the thermal conductivity is calculated by using the Green-Kubo autocorrelation decay method [20, 21]. To ensure the accuracy of our results, the thermal conductivity is averaged over five simulations with different initial velocities. In addition, the size effect [22, 23] is considered and a 400 Å × 400 Å simulation cell containing 38,700 atoms is needed to obtain converged results.

## 3. Results and discussion

The crystal structure of GDY is shown in Figure 1, where the *sp* and *sp*² hybridized carbon atoms are arranged in a hexagonal array. The optimized lattice constants are *a* = *b* = 9.46 Å. The internal coordinates of three carbon atoms marked with $C_1$, $C_2$, and $C_3$ are (0.349, 0.349, 0.000), (0.201, 0.201, 0.000), (0.0708, 0.0708, 0.000), respectively. The coordinates of other fifteen carbon atoms within the unit cell can be derived by utilizing the hexagonal symmetry. The length of carbon-carbon bonds represented by $b_1$, $b_2$, $b_3$, $b_4$ are 1.432, 1.396, 1.233 and 1.339 Å, respectively. Note that the length of $b_1$ and $b_2$ are similar to those found in graphene, while the values of $b_3$ are much smaller due to $sp$ hybridization of carbon atoms. These lattice parameters agree well with those found previously [6], and further confirms the reliability of our calculations. Figure 2 plots the energy band structure of GDY along the high symmetry lines in the hexagonal Brillouin zone. Unlike graphene with a zero gap, we see that GDY is a semiconductor with a direct band gap of 0.48 eV (GGA-PBE value) at the $\Gamma$ point. Such natural gap is believed to be originated from the inhomogeneous $\pi$-bindings between differently hybridized carbon atoms, where the behavior in the chainsaw of GDY are different from those around the carbon hexagons [24, 25]. Moreover, we find that the valence band maximum (VBM) and the conduction band minimum (CBM) are both doubly degenerate, which tends to increase the density-of-state effective mass, as will be discussed later. It should be noted that the standard DFT calculations usually underestimate the band gap when compared to the experimental value. One approach to overcome this deficiency is to calculate the quasiparticle properties with the GW approximation of the many-body effects [26]. Another possibility is using hybrids functionals such as Heyd-Scuseria-Ernzerhof functional (HSE) [27, 28], which also gives a better prediction for the band gap of many semiconductors [29]. In particular, it was previously found [30] that the band gap of graphdiyne is 1.10 eV at the GW level, which is obviously larger than our calculated value at the GGA-PBE level. However, except for such difference in the band gap, we find that the band shape of these two kinds of calculations are almost identical to each other, especially for the energy bands around the Fermi level. As the actual value of band gap does not affect appreciably the thermoelectric properties [31, 32], and both the GW and HSE approaches are very time-consuming and huge computation resources are needed for transport coefficients calculations, we adopt the standard DFT calculations with GGA-PBE in the following discussions.

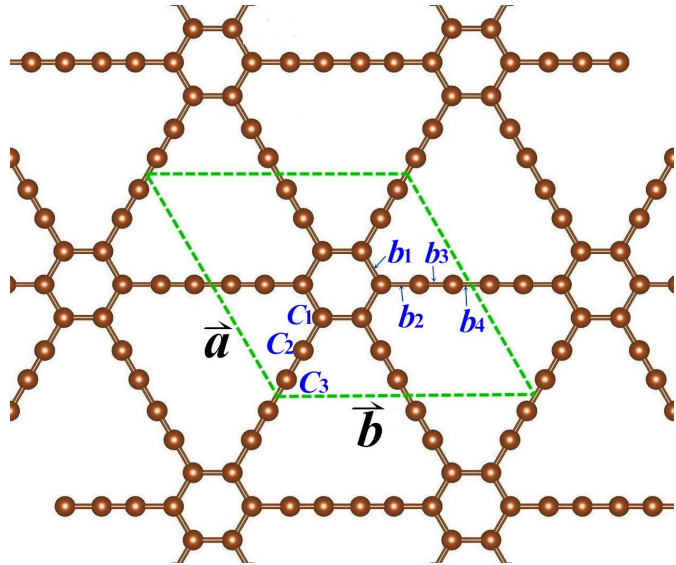

**Figure 1.** The ball-and-stick model of a single-layer GDY. The green dashed line indicates the unit cell with basis vector $\vec{a}$ and $\vec{b}$, where three typical carbon atoms are marked with $C_1$, $C_2$, and $C_3$, and $b_1$, $b_2$, $b_3$, $b_4$ represents different carbon-carbon bonds.

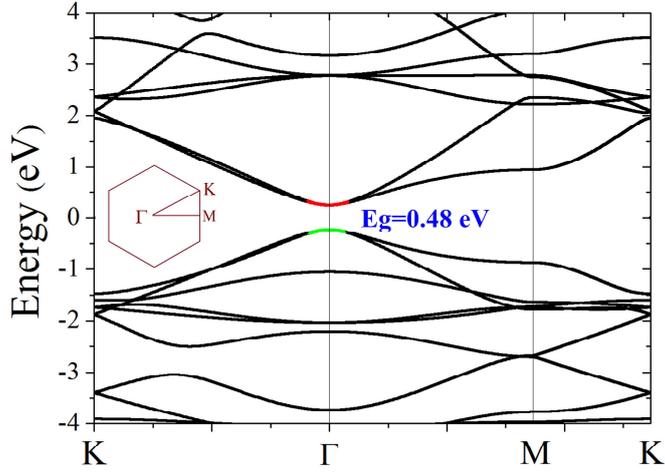

**Figure 2.** Band structure of GDY, where the CBM and VBM are both doubly-degenerate and are red- and green-colored, respectively. The inset shows the hexagonal Brillouin zone.

We now move to the investigations of transport properties. Using the semi-classical Boltzmann theory and relaxation time approximation [17], we can evaluate the electronic transport coefficients from the above-calculated energy band structures. Within this approach, the Seebeck coefficient $S$ is independent of relaxation time $\tau$, while the electrical conductivity $\sigma$ and the power factor $S^2\sigma$ are calculated with respect to $\tau$. Here the relaxation time is predicted from the DP theory [18] considering the acoustic phonons are the main scattering mechanism. For a two-dimensional system the relaxation time at temperature $T$ can be expressed as [33]:

$$\tau = \frac{\mu_c m^*}{e} = \frac{2\hbar^3 C}{3k_B T \left|m^*\right| E_1^2} \tag{1}$$

where $\mu_c$ and $m^*$ are the carrier mobility and effective mass, respectively. Note here a density-of-state effective mass is used for $m^*$, which contains contributions from $k_x$ and $k_y$ directions, and the degeneracy of heavy and light bands at the $\Gamma$ point should be taken into account. The other parameters in Eq. (1) are the elastic constant $C$, the DP constant $E_1$, the unit charge $e$, the reduced Planck constant $\hbar$, and the Boltzmann constant $k_B$. For electrons and holes, we respectively consider the energy shift of CBM and VBM under uniaxial strain. Our calculated results are summarized in Table I. Compared with those of conventional thermoelectric materials, we see that the room temperature relaxation time of GDY is very large for both electrons and holes, which is very beneficial for the electronic transport and highly desirable for good thermoelectric performance. On the other hand, we see from Table I that the electrons exhibit even higher relaxation time than the holes. Since the density-of-state effective mass are similar for electrons and holes, such difference can be attributed to the smaller DP constant of electrons that arisen from the weak scattering by acoustic phonons [34].

**Table I.** The room temperature relaxation time $\tau$, carrier mobility $\mu_c$, DP constant $E_1$, elastic constant $C$, and density-of-state effective mass $m^*$ for GDY.

| Carrier type | $\tau$ (ps) | $\mu_c$ ($10^4$cm$^2$V$^{-1}$s$^{-1}$) | $E_1$ (eV) | $C$ (Jm$^{-2}$) | $m^*$ ($m_0$) |
|---|---|---|---|---|---|
| electrons | 1.16 | 1.35 | 2.66 | 153.97 | 0.151 |
| holes | 0.46 | 0.51 | 4.15 | 153.97 | 0.159 |

Figure 3(a) shows the room temperature Seebeck coefficient $S$ of GDY at different carrier concentration $n$, where positive and negative $n$ indicate the electrons and holes, respectively. Note we are dealing with a two-dimensional system so that the concentration should be understood as the number of carriers per unit area. Around the Fermi level ($n = 0$), we see the Seebeck coefficient $S$ exhibit two obvious peaks, which is $-754\ \mu$V/K at $n = 1.13 \times 10^9\ \text{cm}^{-2}$, and $756\ \mu$V/K at $n = -1.16 \times 10^9\ \text{cm}^{-2}$. The absolute values of these two Seebeck coefficients are much larger than those of most conventional thermoelectric materials, suggesting very favorable thermoelectric performance of GDY. However, we should note that the electrical conductivity $\sigma$ is actually very small at those small carrier concentrations, as shown in Figure 3(b). To maximize the power factor ($S^2\sigma$), one therefore must try to find an optimized carrier concentration, which is shown in Figure 3(c). For $p$-type carriers, we see that the power factor can reach a maximum value of 0.21 Wm$^{-1}$K$^{-2}$ when the carrier concentration is tuned to $-1.5 \times 10^{12}\ \text{cm}^{-2}$. In the case of $n$-type carriers, an even higher power factor of 0.52 Wm$^{-1}$K$^{-2}$ can be achieved. Note that for low-dimensional systems such as the GDY, the definition of cross-sectional area or vacuum thickness has some arbitrariness. In the calculation of band structure of GDY, the supercell geometry has a vacuum thickness of 12 Å. However, to make better comparison with bulk systems (such as three-dimensional graphite), the electrical conductivity $\sigma$ and power factor $S^2\sigma$ of GDY are re-calculated (or converted) with respect to a "more realistic" vacuum thickness of 3.35 Å, which corresponds to the van der Waals distance of graphite and is generally adopted to evaluate the transport coefficients of two-dimensional carbon systems.

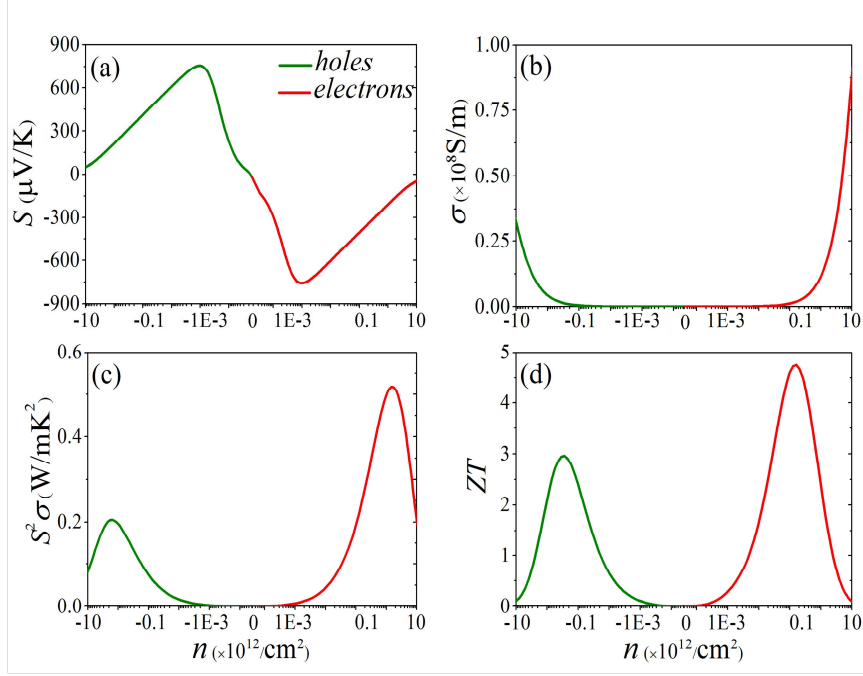

**Figure 3.** The room temperature Seebeck coefficient $S$, electrical conductivity $\sigma$, power factor $S^2\sigma$, and $ZT$ value of GDY as a function of carrier concentration $n$ (using logarithmic coordinates), where the green and red lines correspond to the holes and electrons, respectively.

We next discuss the heat transport coefficients of GDY, which consists of the electronic thermal conductivity $\kappa_e$ and the lattice part $\kappa_l$. In the present work, $\kappa_e$ is calculated by using the Wiedemann-Franz law [35]:

$$\kappa_e = L\sigma T \tag{2}$$

Here the Lorenz number $L$ for two-dimensional systems can be expressed as [36]:

$$L = \left(\frac{k_B}{e}\right)^2 \left[\frac{3F_2}{F_0} - \left(\frac{2F_1}{F_0}\right)^2\right] \tag{3}$$

where $F_i$ is the Fermi integral：

$$F_i = F_i(\eta) = \int_0^\infty \frac{x^i dx}{e^{x-\eta}+1} \tag{4}$$

with $\eta = E_f / k_B T$ is the reduced Fermi energy. The calculated Lorenz number at optimized carrier concentration is listed in Table II.

**Table II.** The optimized room temperature $ZT$ values of GDY. The corresponding carrier concentration $n$, Seebeck coefficient $S$, electronic conductivity $\sigma$, Lorenz number $L$, electronic thermal conductivity $\kappa_e$, and lattice thermal conductivity $\kappa_l$ are also given. The parenthesized values are obtained using GW-corrected band gap, as compared with standard DFT calculations at GGA-PBE level.

| $n$ ($\times 10^{11}$ cm$^{-2}$) | $S$ ($\mu$VK$^{-1}$) | $\sigma$ ($\times 10^6$ Sm$^{-1}$) | $L$ ($\times 10^{-8}$ V$^2$K$^{-2}$) | $k_e$ (Wm$^{-1}$K$^{-1}$) | $k_l$ (Wm$^{-1}$K$^{-1}$) | $ZT$ |
|---|---|---|---|---|---|---|
| −2.74 | 323.90 | 1.20 | 1.52 | 5.46 | 7.30 | 3.0 |
| (−3.14) | (323.94) | (1.20) | (1.52) | (5.48) | | (3.0) |
| 1.62 | −363.46 | 1.91 | 1.51 | 8.62 | 7.30 | 4.8 |
| (1.59) | (−379.21) | (1.64) | (1.50) | (7.38) | | (4.8) |

For the phonon transport of GDY, we use the EMD method where the lattice thermal conductivity $\kappa_l$ can be obtained by using the Green-Kubo relation [20, 21]:

$$\kappa_l = \frac{1}{V k_B T^2} \int_0^{\tau_m} \langle J(\tau) J(0) \rangle d\tau \qquad (5)$$

Here $\langle J(\tau) J(0) \rangle$ is the heat current autocorrelation function, $V$ is the system volume, $k_B$ is the Boltzmann constant, and $T$ is the system temperature. As done for the electronic transport coefficients, we use the same vacuum thickness of 3.35 Å to evaluate the lattice thermal conductivity of a two-dimensional system so that the $ZT$ value does not depend on the arbitrary definition of the cross-sectional area or vacuum thickness. In another word, if we use a different vacuum thickness (say, the initial value of 12 Å), there will be changes of individual transport coefficients ($\sigma$, $\kappa_e$, $\kappa_l$), while the $ZT$ value remains the same since the effect of arbitrariness is cancelled by the numerator and denominator of the $ZT$ formula. Our EMD simulation gives a lattice thermal conductivity of 7.30 Wm$^{-1}$K$^{-1}$ for GDY, which is much lower than those of other carbon allotropes such as graphene and carbon nanotubes, suggesting very favorable thermoelectric applications of GDY. The reduced thermal conductivity of GDY is believed to be caused by the presence of acetylenic linkages ($sp$ carbon bond), which gives rise to inefficient heat transfer by lattice vibrations [37] compared with the $sp^2$ bonded carbon materials. Table II summarizes all the transport coefficients discussed above, from which we can now evaluate the thermoelectric performance of GDY. Figure 3(d) shows the room temperature $ZT$ values as a function of carrier concentration. At optimized concentration, we see that the $ZT$ value of GDY can be reached to 3.0 for $p$-type carriers and 4.8 for $n$-type carriers. These $ZT$ values significantly exceed most laboratory results reported so far, and are either equal to or larger than the target value ($ZT$=3.0) for the commercial applications of thermoelectric materials. We want to emphasize that the above-calculated transport coefficients and $ZT$ values are based on the GGA-PBE band gap of 0.48 eV. If the more sophisticated GW technique is used, one can obtain an accurate band gap of 1.10 eV [30]. To double-check that the actual value of band gap will not appreciably change the thermoelectric properties [31, 32], we make a rigid shift of the GGA-PBE band structure so that the band gap is increased to the GW value of 1.10 eV. At such accurate band gap, we find there are indeed very small changes of the optimized electronic transport coefficients (see the parenthesized values in Table II). The reason is that although increasing band gap may change specific value of transport coefficients (for example, increasing the peak value of Seebeck coefficient), the transport coefficients at the optimized concentration will not be changed substantially. In fact, we see from Table II that the GGA-PBE and GW approaches give identical $ZT$ values for both the $n$-type and $p$-type GDY. In this regard, we believe the present calculations give a reasonable good prediction of the thermoelectric properties of GDY.

## 4. Summary

Using a multiscale approach combining first-principles, Boltzmann theory, and molecular dynamics simulations, we show that the recently fabricated GDY could be optimized to exhibit very high figure of merit. Our work not only provides a materials-specific system to realize the concept of PGEC, but also gives strong evidence that good thermoelectric performance can be achieved in previously unexpected carbon materials. It should be mentioned that single-layer GDY is currently only available as fractions in experiments [5], and great efforts should be devoted to the possible realization of this environmentally-friendly thermoelectric materials containing earth-abundant carbon element.

## Acknowledgements

We thank financial support from the National Natural Science Foundation (Grant

No. 51172167 and J1210061) and the "973 Program" of China (Grant No. 2013CB632502).